\documentclass{article}

\usepackage{PRIMEarxiv}

\usepackage[utf8]{inputenc} 
\usepackage[T1]{fontenc}    
\usepackage{hyperref}       
\usepackage{url}            
\usepackage{booktabs}       
\usepackage{amsfonts}       
\usepackage{nicefrac}       
\usepackage{microtype}      
\usepackage{lipsum}
\usepackage{fancyhdr}       
\usepackage{graphicx}       
\usepackage{amsmath}
\usepackage{rotating}
\usepackage[title]{appendix}%

\graphicspath{{media/}}     

\title{Digital Skills Formation in Gendered Peer Networks: Exploring advice giving and taking in classrooms.
}

\author{
  Petro Tolochko \\
  University of Vienna \\
  Department of Communication \\
  Vienna, Austria\\
  \texttt{petro.tolochko@univie.ac.at} \\
  \And
  Jana Bernhard-Harrer \\
  University of Vienna \\
  Department of Communication \\
  Vienna, Austria\\
  \And
  Azade E. Kakavand \\
  University of Vienna \\
  Department of Communication \\
  Vienna, Austria\\
  \And
  Aytalina Kulichkina \\
  University of Vienna \\
  Department of Communication \\
  Vienna, Austria\\
  \And
   Hyunjin Song \\
  Yonsei University \\
  Department of Communication \\
  Seoul, South Korea\\
  \And
  Hajo G. Boomgaarden \\
  University of Vienna \\
  Department of Communication \\
  Vienna, Austria\\
}

\begin{document}
\maketitle

\begin{abstract}
The digitalisation of childhood underscores the importance of early digital skill development. To understand how peer relationships shape this process, we draw on unique sociocentric network data from students in classrooms across three countries,  focusing on peer-to-peer advice-giving and advice-seeking networks related to digital skills. Using exponential random graph models, we find that digital skills systematically spread through peer interactions: higher-skilled students are more likely to be sought for advice while less likely to seek it themselves. Students perceived as highly skilled are more likely to seek and offer advice, but it has limited influence on being sought out by others. Gender plays a significant role: girls both seek and give more advice, with strong gender homophily shaping these interactions. We suggest that digital skills education should leverage the potential of peer learning within formal education and consider how such approaches can address persistent divides.
\end{abstract}


\section{Introduction}

In increasingly digitized societies, digital skills have become central to successfully navigating personal, civic, and professional life. Acquiring or possessing digital skills is associated with individuals' well-being \cite{livingstone_outcomes_2023, vissenberg_digital_2022}, educational goal attainment \cite{pagani_impact_2016, mehrvarz_mediating_2021}, and a favorable positioning on the labor market \cite{kiss_digital_2017, sanchez-canut_professional_2023}. Importantly, children's and adolescents' ability to cope with risky or negative online experiences and prevent further harm depends critically on their digital skills \cite{vandoninck_digital_2010, quandt_dark_2022}. To prepare young people for life in a digitized world, it is indispensable to prioritize the development of their digital skills \cite{wendt_social_2023, jackman_addressing_2021}, as technology affects childhood and adolescence more than ever before \cite{danby_situated_2018}. However, despite the increasing ubiquity of digital technology, a quarter of students worldwide only hold basic digital skills \cite{fraillon_international_2023}.

Knowing how to navigate and use digital tools from an early age has implications beyond the individual, as it serves as a foundation for reducing digital inequalities for generations ahead \cite{van_deursen_internet_2011, liu_social_2022, van_de_werfhorst_digital_2022}. As digital technology becomes essential in daily life, digital skills are increasingly vital. However, marginalized individuals, particularly those at the intersection of multiple identities, such as gender or socio-economic status (SES), still face compounded barriers to acquiring digital skills, reinforcing systemic inequalities further. Young people's disadvantages in access to education or training and the internet systematically cause digital skill gaps \cite{noauthor_decoding_2024}. Despite some progress along the gender dimension, with female students showing high levels of computer and information literacy during school years in some contexts \cite{fraillon_international_2023}, adult women still face persistent disadvantages in income or representation in high-level ICT-related occupations worldwide \cite{perez-felkner_computing_2024}. This leaky pipeline \cite{yang_holistic_2021} underlines the importance of a gendered perspective on digital skills acquisition.

Given the centrality of digital skills, empirical studies have engaged with the question of why some children and adolescents have higher digital competence than others \cite{haddon_childrens_2020}. This body of work has primarily focused on individual characteristics and capacities rather than the socio-contextual aspects that may contribute to children and adolescents holding digital skills. However, a social structural perspective on digital skills acquisition argues that young people often acquire digital skills within their immediate environments such as schools, families, and particularly within their peer networks \cite{mehrvarz_mediating_2021} -- and highlights the importance of networks and spaces in shaping engagement with digital technologies \cite{helsper_youth_2020}. Given the well-established influence of peer networks on children's and adolescents' development \cite{jenks_constructing_2004, karuovic_use_2016, steinsbekk_new_2024}, this study introduces and systematically tests the relationship between peer networks and digital skills acquisition. Specifically, we investigate the role of gender and digital skill levels in such networks as potentially decisive factors in understanding digital skill inequalities. How do digital skills affect skill-sharing networks, and what role do gender and prior digital skills play in their formation? The findings contribute to public discourse on children's digital skills and offer practical implications for educators and policymakers.

Adolescence is a time in which strong ties with peers are developed and maintained \cite{steinsbekk_new_2024}, and such relationships form an essential part of adolescents' daily interactions \cite{bronfenbrenner_bioecological_2007}. Renzini and colleagues \cite{renzini_status_2024} argue that understanding the formation of social networks requires considering the context-specific micro-level interplay of network dynamics and individual preferences. It was shown that reciprocity (i.e., a tendency for relationships to be mutual), popularity (i.e., some individuals become highly connected in a network), and triadic closure (i.e., if two people share a mutual friend, they are more likely to become friends themselves) were affecting the formation of preschool children's networks \cite{schaefer_fundamental_2010}. Research also indicates that adolescents tend to form networks based on similarity, driven by homophilic social selection (i.e. a tendency for people to form connections with others who are similar to them) and social influence \cite{montgomery_improving_1977}. These networks become more exclusive with age, and females are generally more connected than males \cite{urberg_structure_1995}. Boomgaarden and colleagues \cite{boomgaarden_report_2022} find that homophily in classroom friendship networks is strongly influenced by demographic factors such as gender. Peer relationships in early adolescence are moderately stable, with tendencies towards reciprocity, network closure, and gender similarity shaping network evolution \cite{lubbers_dynamics_2011}.

Regarding the development of digital skills, the existing literature provides insight into different types of social structures, for example, by looking at the impact of factors related to family contexts, especially their socioeconomic status \cite{livingstone_balancing_2010, van_de_werfhorst_digital_2022}, and patterns of parental mediation \cite{vandoninck_digital_2010, senkbeil_how_2023, perez-sanchez_effects_2022, kalmus_towards_2022}. In addition to family structures, the focus is generally on the impact of formal schooling \cite{wang_classroom_2020, cantin_change_2004, aesaert_contribution_2015, beilmann_role_2022}, distinguishing the role of teachers and the integration of digital technology both as hardware and within curricula \cite{aesaert_contribution_2015}. Yet, there are good reasons to assume that digital skills are also -- and maybe even more centrally -- acquired through informal social relationships \cite{mehrvarz_mediating_2021, mills_coding_2024}. Unfortunately, this perspective has been largely ignored in the literature on digital skill acquisition of children and adolescents so far, along with a lack of systematic engagement with the role of peer networks aside from their relationship with academic success or psychological well-being (see \cite{wang_classroom_2020} for a systematic review).

Students' collaborative learning can be conceptualized as a network in which those who work together are connected, with ties strengthening as interactions become more frequent. Children can acquire knowledge by seeking help from peers, but research also highlights tutor effects, suggesting that students can learn through the act of advising others \cite{webb_peer_2010, ryabov_adolescent_2011, fujiyama_peer_2021}. In our study, we thus differentiate between the acts of giving advice to and seeking advice from friends regarding digital skills, taking into account that peer learning can take different forms. We seek to understand how digital skills shape these forms of peer learning. Skill-sharing networks are shaped by patterns of advice-seeking and advice-giving, which are influenced by complex interactions among individual characteristics, skill levels, and social dynamics. To explore these mechanisms, we analyze the role of digital skills in structuring advice networks. Specifically, we ask:\newline

\noindent \textbf{RQ1a}: Do higher levels of digital skills increase the likelihood that students will give advice and be sought for advice by others?\newline

\noindent \textbf{RQ1b}: Does being perceived as having strong higher skills increase the likelihood that students will give advice and be sought for advice by others?\newline

\noindent \textbf{RQ1c}: Do students with high digital skills share advice more frequently with others who also have high digital skills?\newline

Such informal peer networks are likely to be gender-structured \cite{boomgaarden_report_2022}, similar to civic resource acquisition \cite{djupe_present_2007} and social interactions in general \cite{djupe_political_2018}. However, peer networks in digital skills acquisition and the role of gender in these networks have not received systematic attention. Haddon et al.'s \cite{haddon_childrens_2020} literature review shows mixed findings regarding the relationship between gender and digital skills. Studies using performance-based measures often find no significant gender differences in girls' and boys' digital skill levels, or even that girls outperform boys. In contrast, studies relying on self-reports tend to find that boys rate their digital skills higher than girls. Similar patterns emerged during the COVID-19 pandemic, with home and distance learning becoming important forms of education \cite{peng_cognition_2022, goudeau_sebastien_why_2021, jackman_addressing_2021}. Looking more broadly at the intersection of digital skills, knowledge, and attitudes, Campos and Scherer \cite{campos_digital_2024} show that girls demonstrate superior digital knowledge and skills compared to boys, with gender disparities shaped by variances in attitudes towards technology and socioeconomic factors. Recently, research has taken the much needed step to go beyond a binary notion of gender, showing that non-binary children and adolescents tend to have more experience creating digital content and using the Internet to access information on mental and physical health \cite{de-coninck_gendered_2023}. In sum, gender affects self-assessments of digital skills: girls tend to rate themselves lower than boys, despite performance-based evidence showing little to no gender difference \cite{haddon_childrens_2020}. Moreover, this self-reported gap tends to widen when children report experiences with discrimination \cite{mascheroni_explaining_2022}. In our study, we are the first to explore how gender relates to skill-sharing relationships in peer networks and how this may contribute to structural inequalities in digital skill acquisition. Being aware of gender strongly structuring peer networks in adolescence generally \cite{lubbers_dynamics_2011, ko_interplay_2015}, our study stresses how gender-based inclusion and exclusion can lead to inequalities in digital skills attainment. Since skill-sharing is likely driven by perceived -- rather than actual -- competence, and girls tend to underrate their digital skills, we expect self-perceptions to play a key role in shaping these dynamics. 

Specifically, we focus on how gender can affect digital skill-sharing, advice-giving, and advice-seeking networks. First, we seek to better understand whether girls are more likely to give advice or seek advice than boys in these networks. Therefore, we ask:\newline

\noindent \textbf{RQ2a}: Are girls more likely to take on advice-seeking roles in peer networks than boys?\newline

\noindent \textbf{RQ2b}: Are girls more likely to take on advice-giving roles in peer networks than boys?\newline

Building on prior literature on digital skills and peer networks, we further hypothesise that gender influences how students choose their advice partners. Based on the observation that stereotypes may affect the skill levels perceived by oneself and others \cite{haddon_childrens_2020}, we assume that girls are less likely to be the receiver of advice-seeking and more likely to receive advice.\newline

\noindent \textbf{H1a}: Students are more likely to seek advice from boys than girls.\newline

\noindent \textbf{H1b}: Students are more likely to give advice to girls than boys.\newline

Furthermore, provided previous findings about strong gender homophily in peer networks of children and adolescents \cite{boomgaarden_report_2022}, we assume that this general tendency should also affect skill-sharing relationships.\newline

\noindent \textbf{H2a}: Students are more likely to seek advice from peers of the same gender.\newline

\noindent \textbf{H2b}: Students are more likely to give advice to peers of the same gender.\newline

\begin{figure}[t]
    \centering
        \includegraphics[width=\linewidth]{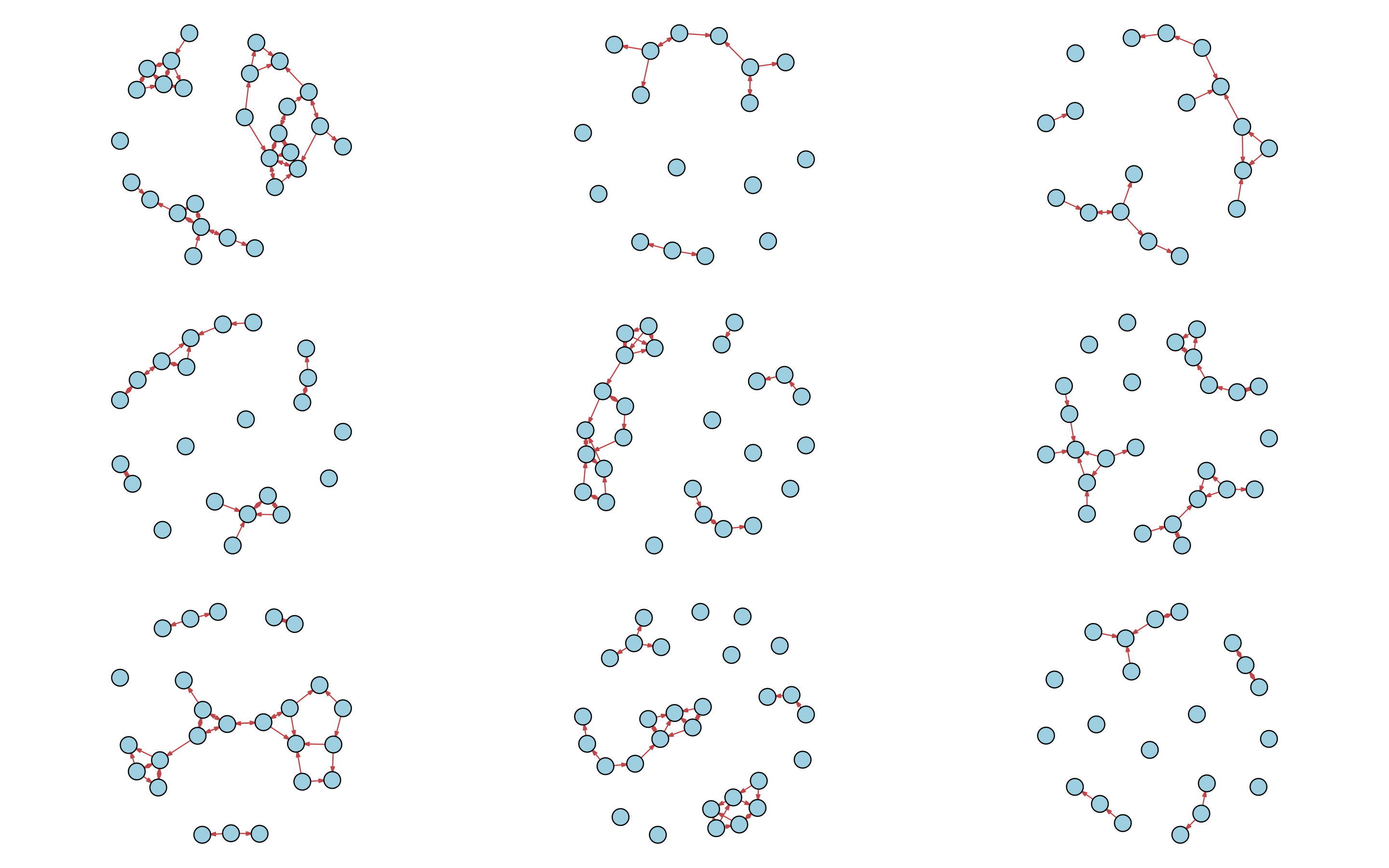}
        \caption{A Sample of 9 Directed Advice Giving Networks.}
    \label{fig:networks}
\end{figure}

These Research Questions and Hypotheses are addressed through the use of a unique dataset that captures 372 sociocentric networks in 18 schools across three countries (Germany, Italy, and Portugal) at different points in time \cite{machackova2024digital} (a visualisation of 9 networks from the data is provided in Figure~\ref{fig:networks}). These sociocentric network data allow a robust estimation of the social context through which digital skills are shared. Although the importance of digital skills has led to much scholarly attention, there is no clear conceptualization of digital skills that is shared beyond fields, and thus, also the operationalization of what constitutes digital skills varies greatly between individual studies \cite{haddon_childrens_2020, audrin2022key}. Our conceptualization and measurement of digital skills follow the ySKILLS projects' approach \cite{helsper_youth_2020}, differentiating five dimensions of digital skills in the youth Digital Skills Indicator (yDSI): (1) technical and operational skills, (2) programming skills, (3) information navigation and processing skills, (4) communication and interaction skills, and (5) content creation and production skills.

\section{Methods}

\subsection{Data}

The data were collected as part of collected as part of an international Horizon 2020-funded project ySKILLS \cite{machackova2024digital}. The network data was obtained in three waves from Spring 2021 to Spring 2023, and comes from three different countries -- Germany (N = 127), Portugal (N = 128), and Italy (N = 117), 372 networks in total. When collecting network data, students (egos) were first asked to select three of their best friends in the class (alters), and then, out of these three alters who they ask for advice most often and to whom they give advice most often, thus creating two separate directed network types -- with advice seeking and advice giving ties. The average distribution of edges was (M = 20.77; SD = 9.55) and (M = 22.25; SD = 10.24) for seeking and giving networks, respectively. The average student age was (M = 15.43; SD = 1.36) years, with roughly 50\% of students being female. The dataset, therefore, consists of multiple sociocentric (known boundary) networks, providing a unique opportunity to understand general patterns across multiple networks.

Several networks contained missing values on the nodal level. Missing data were imputed using the multivariate imputation by chained equations (MICE) method \cite{azur2011multiple, van2011mice}. This method deals with missing data by iteratively imputing each missing value multiple times using regression models, conditioned on the other variables. In each iteration, MICE fills in missing values for each variable one at a time, using the available observed values of other variables in a chain of sequential conditional models, until the imputations converge. The imputation was done on the complete dataset, before individual networks were extracted.

\subsection{Digital Skills Variables}

\textit{Overall Digital Skills}. A constructed a composite measure of overall digital skills that captures students' overall digital competence was used. This item integrates responses from multiple subdimensions measured within the scope of the project, including technical skills, information processing skills, programming skills, content creation skills, and communication skills. The overall digital skills variable was thus made up of 21 separate questions (relating to various subdimensions of digital skills). All items showed strong internal consistency across domains, and the composite score was validated. Please refer to \cite{machackova2024digital} and \cite{helsper2020youth} for detailed information on data collection, scale creation, and validation. The item was scaled to range from 0 to 1 (M = 0.45; SD = 0.1).

\textit{Perceived Digital Skills}. The perceived digital skills variable was measured by a single item: \textit{How good do you think [Friend] is at using the internet and technologies such as phones?}. The item was measured using a 5-point Likert scale ranging from 1 (not good at all) to 5 (excellent). The combined perceived skills score was calculated by averaging the incoming scores on this item from alters to the ego. For example, if A perceives X to be 2, B perceives X to be 3, and C perceives X to be 5, the perceived digital score of X is 3.33. The item was scaled to range from 0 to 1 (M = 0.46; SD = 1.66). Example items measuring various subdimesions of digital skills are presented in the Appendix.

\subsection{Method and Models}

We used a hierarchical meta-analysis estimation approach described by \cite[]{tolochko2024same}. Using this methodology, each individual network is estimated using an Exponential Random Graph Model \cite[]{lusher2013exponential}. After ERGMs have been estimated, the coefficients and their standard errors are used as new data points for a Bayesian hierarchical model, which allows to pool the estimates together, resulting in a conservative estimation of effects. This approach allows treating each individual social network as a draw from a population distribution, instead of relying on a single network to draw conclusions about the population.

\subsubsection{Hierarchical Bayesian Model}

Formally, we model the network data as follows:

\begin{samepage}
\begin{align*}
\hat{\theta}_k &\sim \mathcal{N}(\theta_k, \sigma_k^2) \\
\theta_k &\sim \mathcal{N}(\mu, \tau_k) \\
\mu &\sim \mathcal{N}(\alpha_{\text{country}}, 1) \\
\alpha_{\text{country}} &\sim \mathcal{N}(0, \tau_{\text{country}}) \\
\tau_k &\sim \text{HalfCauchy}(0, 1) \\
\tau_{\text{country}} &\sim \text{HalfCauchy}(0, 1)
\end{align*}
\end{samepage}

\noindent where $\hat{\theta}_k$ is the estimated ERGM effect size of network $k$, $\sigma^2_k$ is the variance of the estimate, $\theta_k$ is the true effect size drawn from a Normal distribution with mean $\mu$ (pooled effect size across networks) and variance $\tau_k$ (modeling between network heterogeneity). Finally, $\alpha_{country}$ and $\tau_{country}$ are hierarchical additions to account for pooled country effects and country heterogeneity, respectively.

\subsection{Exponential Random Graph Model Specifications}

For our research questions and hypotheses, we model each network with the following specifications (before pooling the estimates). The standard ERGM specification is as follows:

\[
P(Y = y) = \frac{1}{\kappa(\boldsymbol{\theta})} \exp\left( \boldsymbol{\theta}^\top \cdot \boldsymbol{g}(y) \right)
\]

\noindent where $\boldsymbol{\theta}$ is the parameter vector and $\boldsymbol{g(y)}$ is vector of ``sufficient network statistics'' (structural terms and nodal covariates). We further present $\boldsymbol{{g}(y)}$ for each of the RQs and Hypotheses.

\subsubsection{Research Question 1}

\[
\boldsymbol{g}_{\text{RQ1}}(y) = 
\begin{bmatrix}
\text{edges}(y) \\
\text{mutual}(y) \\
\text{nodeocov}(\text{skills}, y) \\
\text{nodeicov}(\text{skills}, y) \\
\text{nodeocov}(\text{perceived skills}, y) \\
\text{nodeicov}(\text{perceived skills}, y) \\
\text{absdiff}(\text{skills}, y)
\end{bmatrix}
\]

\noindent \textbf{RQ1a}: Objective Digital Skills effects \textit{nodeocov(skills)} and \textit{nodeicov(skills)}.

\noindent \textbf{RQ1b}: Perceived Digital Skills effects \textit{nodeocov(perceived skills)} and \textit{nodeicov(perceived skills)}.

\noindent \textbf{RQ1c}: Homophily effect \textit{absdiff(skills)}.

\subsubsection{Research Question 2}

\[
\boldsymbol{g}_{\text{RQ2}}(y) = 
\begin{bmatrix}
\text{edges}(y) \\
\text{mutual}(y) \\
\text{nodeocov}(\text{skills}, y) \\
\text{nodeicov}(\text{skills}, y) \\
\text{nodeocov}(\text{perceived\_skills}, y) \\
\text{nodeicov}(\text{perceived\_skills}, y) \\
\text{absdiff}(\text{skills}, y) \\
\text{nodeocov}(\text{female}, y)
\end{bmatrix}
\]

\subsubsection{Hypothesis 1: Receiver Effects of Gender}

\[
\boldsymbol{g}_{\text{H1}}(y) = 
\begin{bmatrix}
\text{edges}(y) \\
\text{mutual}(y) \\
\text{nodeicov}(\text{female}, y)
\end{bmatrix}
\]

\subsubsection{Hypothesis 2: Gender Homophily}

\[
\boldsymbol{g}_{\text{H2}}(y) = 
\begin{bmatrix}
\text{edges}(y) \\
\text{mutual}(y) \\
\text{nodematch}(\text{female}, y)
\end{bmatrix}
\]

\subsubsection{Network Terms}

\begin{itemize}
    \item \textit{edges}: this term captures the overall propensity for tie formation in a network, and serves as a baseline (similar to an intercept) parameter.
  \item \textit{mutual}: this term models reciprocity, indicating whether directed ties are more likely to be reciprocated.
  \item \textit{nodeocov(X)}: Outdegree covariate effect; models whether nodes with higher values on variable \textit{X} are more likely to receive ties.
    \item \textit{nodiocov(X)}: Indegree covariate effect; models whether nodes with higher values on variable \textit{X} are more likely to send ties.
    \item \textit{absdiff(X)}: Attribute Homophily (Absolute Difference); models whether nodes are more likely to form ties with others who have similar values on variable \textit{X}.
    \item \textit{nodematch(X)}: Attribute Homophily (Categorical Attribute); models homophily based on a categorical nodal attribute \textit{X}, such as gender, or group membership. It measures whether ties are more likely to form between nodes that share the same value on variable \textit{X}.

\end{itemize}

\section{Results}\label{sec2}

Our models explain the presence of a digital skill sharing relationship (tie) between two students (nodes) in a classroom by differentiating between advice-giving and advice-seeking networks. In an advice‐giving network, the model shows how a student's characteristics (e.g., high digital skills or being perceived as highly skilled) increase the likelihood of offering digital skills advice to peers. Conversely, in an advice‐seeking network, the same nodal attributes may influence how often a student is approached for advice on digital skills. These two networks are not just the opposite sides of the same coin, as the reasons for giving advice might very well differ from those for seeking advice from others. 

Four different nodal effects are modeled. First, the out-degree (sender) covariate effect captures how a node's own attribute affects its propensity to initiate ties. For example, in the advice-seeking network, the student's gender might increase the tendency to seek advice, implying that a certain gender is more proactive in asking for help with digital skills. Second, the in-degree (receiver) effect shows how a nodal attribute influences the likelihood of receiving ties. Third, in models assessing same-gender preferences, the nodematch term is used to capture the tendency for students of the same gender to form advice ties, directly reflecting homophily based on gender. Finally, the absolute difference term measures the effect of similarity between students on tie formation, capturing homophily based on continuous values. In addition to the nodal covariates, there are two structural network effects -- density and reciprocity. The density parameter represents the baseline probability of forming any tie in the network, and the reciprocity effect indicates that when one student offers advice to another, there is an increased likelihood that advice will be reciprocated.

Beginning with the first research question (RQ1a) looking at the advice-seeking networks, we find that students with higher total digital skills are less likely to seek advice than those with lower levels, shown by the negative sender effect (out-degree) for total skills  ($\beta=-0.95$, 95\% CI [-1.08; -0.83]). Students with higher levels of skills are more likely to be sought for advice, indicated by the positive receiver (in-degree) effect ($\beta=0.47$, 95\% CI [0.35; 0.60]). In a similar way, in the advice-giving networks, the sender effect is positive ($\beta=0.16$, 95\% CI [0.04; 0.28]), and the receiver effect is negative ($\beta=-0.34$, 95\% CI [-0.49; -0.25]), suggesting that students with higher skills tend to give more and receive less advice from others.

For the perceived digital skills (RQ1b), we find that, in the advice-seeking networks, there is a positive sender effect ($\beta=0.12$, 95\% CI [-0.11; -0.13]), showing that people who are perceived to be highly digitally skilled are more likely to seek advice from others. The receiver effect is null, suggesting no effect of perceived digital skills on being approached by others for advice. Within the advice-giving networks, the sender effect is small but positive ($\beta=0.07$, 95\% CI [0.06; 0.08]), meaning a higher perception of digital skills by others increases advice-giving behaviour. Likewise, there is a small yet positive receiver effect ($\beta=0.02$, 95\% CI [0.01; 0.03]), indicating that a higher perception of skills slightly increases the chances of receiving advice from others. This pattern suggests that lower perceived skills may be associated with lower overall activity in the network, potentially reflecting that those who are less actively seeking or giving advice are perceived to have lower skills than those who manifest active behavior.

Finally, for RQ1c, we look at the absolute difference between the total digital skills. The coefficient is negative for both the advice-seeking and advice-giving networks ($\beta=-0.42$, 95\% CI [-0.55; -0.30]) and  ($\beta=-0.37$, 95\% CI [-0.49; -0.25]), respectively. This indicates that the chance of forming a tie substantially decreases as the discrepancy between the skills of two students increases (i.e. when the skills become more dissimilar). Therefore, students tend to form advice ties with peers who have more similar levels of digital proficiency. This partially supports the previous finding, indicating that perceived digital skills do not affect the likelihood of being approached for advice. Overall, addressing RQ1c, we thus find that students who report strong digital skills do share advice with one another.

For the second set of RQs and Hypotheses, we add an additional parameter -- a sender effect for being female. In advice-seeking networks (RQ2a), there is a positive sender effect of being female ($\beta=0.18$, 95\% CI [0.13; 0.24]), suggesting that girls are more likely to seek advice. Similarly, in the advice-giving networks, the effect is also positive ($\beta=0.06$, 95\% CI [0.01; 0.11]) but smaller in magnitude, indicating that while girls are also more likely to give advice, the gender effect is less pronounced in advice-giving compared to advice-seeking behaviour.

In order to test both hypotheses, we have estimated a simplified model with sender and homophily effects of gender -- that is, a model that examines how a student's own gender influences their likelihood of initiating advice ties (sender effect) and the tendency for students to form advice relationships with peers of the same gender (homophily effect). In the advice-seeking networks, the receiver effect for gender (H1a) was positive ($\beta=0.08$, 95\% CI [0.02; 0.13]), indicating that students are more likely to seek advice from girls rather than boys, therefore, not supporting H1a. In the advice-giving networks, the receiver effect of gender (H1b) is also positive ($\beta=0.12$, 95\% CI [0.07; 0.17]), suggesting that students are more likely to direct their advice toward girls, supporting H1b.

Moving to H2, the gender homophily effect is strong and positive in both the advice-seeking and advice-giving networks ($\beta=1.13$, 95\% CI [1.07; 1.20] for seeking and $ \beta=1.11$, 95\% CI [1.04; 1.18] for giving, respectively). These results indicate that students prefer to establish both advice-seeking and advice-giving ties with students of the same gender, supporting H2a and H2b. In relation to gender, we thus find overall that students tend to both seek and offer advice within same-gender peer groups. 

\begin{figure}[t]
    \centering
        \includegraphics[width=\linewidth]{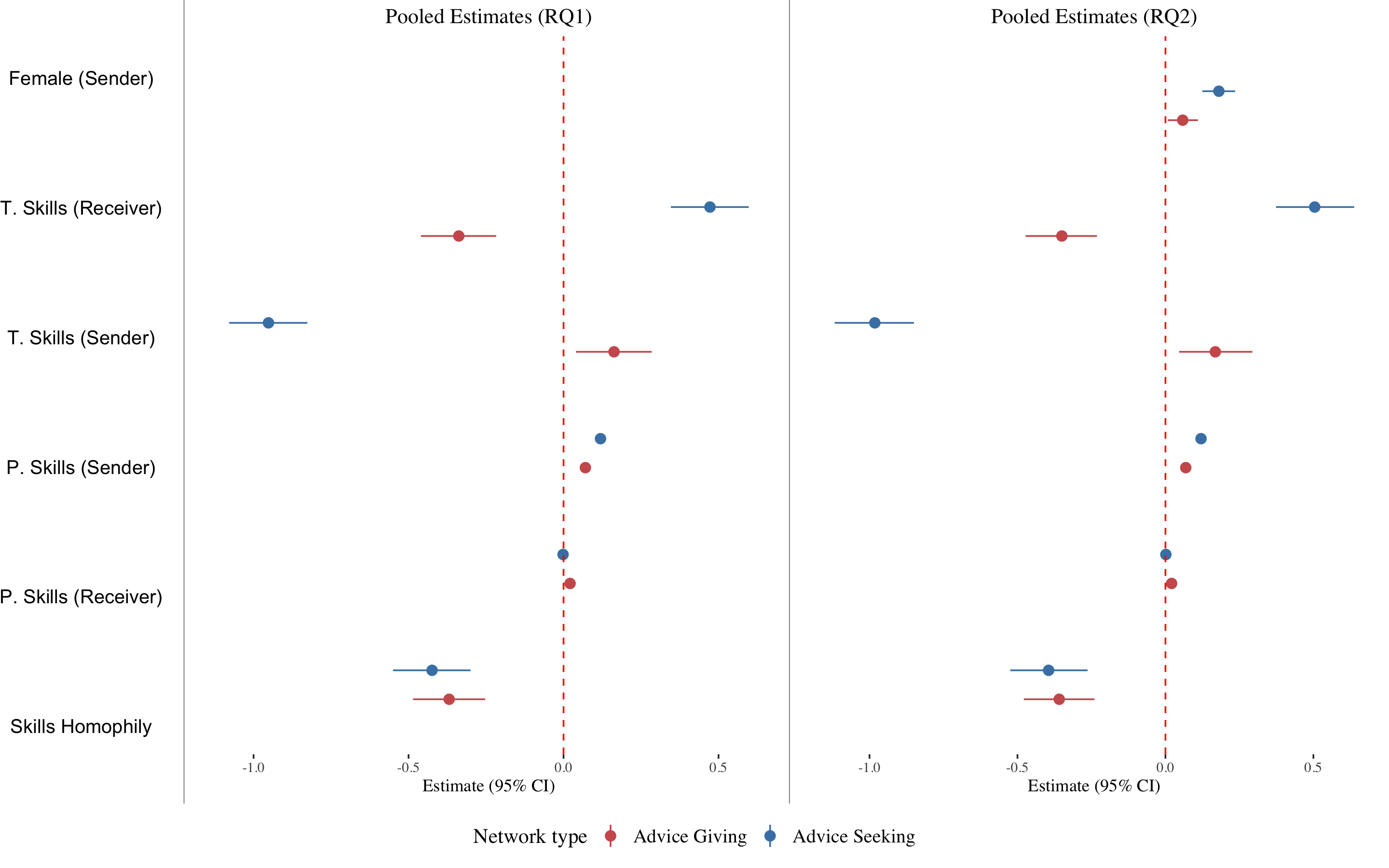} 
        \caption{Posterior Distributions of Pooled ERGM estimates. Left Pane: RQ1 model, Right Pane: RQ2 model.\\
        Note: Density and Reciprocity parameters were excluded from the Figure to improve readability. The full specification is present in Table \ref{tab:appendix} in the Appendix. ``T. Skills": "Total Skills"; ``P. Skills": ``Perceived Skills''.}
    \label{fig:estimates}
\end{figure}

Figure \ref{fig:estimates} shows the estimated posterior distributions of network statistics for RQ models.

\section{Discussion and Conclusion}\label{sec3}

Aiming to better understand the importance of peer networks in digital skills learning and to assess whether gender systematically structures these networks, our research provides novel insights into how adolescents engage in digital skill-sharing. Our findings show that students with higher self-reported digital skills are less likely to seek advice, but more likely to be sought out and to give advice. However, in terms of perceived digital skills, students who are perceived as highly skilled are more likely to seek and give advice, but are not actively being sought out. This may suggest that their visible involvement increases perceptions of expertise, but may also discourage others from seeking them out due to perceived inaccessibility or hesitancy. In terms of peer selection, students are more likely to form advising relationships with others who have similar skill levels. Gender also plays a significant role, and strong gender homophily suggests that advice networks are primarily shaped by same-sex interactions. Girls are more likely than boys to seek and give advice, but more so when seeking advice. Students are also more likely to seek advice from and give advice to girls.

By focusing on peer networks, our study confirms that digital skills are actively shared within adolescents' informal social circles -- a finding that supports and extends previous assumptions about how learning occurs in these contexts. Our approach highlights the value of examining collaborative learning through the lens of network structures, echoing arguments that everyday interactions shaped by micro-level dynamics play a central role in skill acquisition \cite{bronfenbrenner_bioecological_2007, webb_peer_2010}. While previous research has emphasized the role of formal education and family contexts \cite{aesaert_contribution_2015, livingstone_balancing_2010}, our study foregrounds the importance of informal peer networks as a critical and underexplored site for digital skill development.

Our findings extend the digital literacy literature, which has traditionally focused on individual-level explanations of competence, by emphasizing the social contexts and relational mechanisms through which digital skills are developed and recognized \cite{haddon_childrens_2020}. Through the network approach, we show how both objectively measured and peer-perceived skills shape advice-seeking and giving behaviours. While objectively measured digital skills are strongly associated with advice-seeking and giving, perceived digital skills have weaker and less consistent effects. This divergence underscores the importance of considering both actual skill levels and social recognition when studying digital learning, especially since peer nominations may reflect visibility, confidence, or bias rather than actual skill. Future research could examine possible antecedents of digital skill recognition among adolescents, as they may reflect structural inequalities and influence children's opportunities for advancement. Also, it would be desirable to incorporate skill performance measures into such large-scale data collections.

The tendency to form bonds with peers with similar skill levels further supports the role of homophily in structuring informal learning. Previous research has shown that adolescent networks tend to form along shared attributes such as gender or interests \cite{montgomery_improving_1977, steinsbekk_new_2024}, but our study highlights that functional homophily based on skills also guides who learns from whom. While this can foster effective communication and collaboration, it can also have unintended consequences: when high- and low-skilled students remain in separate peer clusters, opportunities for informal digital learning are unevenly distributed. As a result, lower-skilled students may have fewer opportunities to access peer support, reinforcing digital inequality.

Gender further structures these learning dynamics. Our findings show that girls are more engaged in both the advice-seeking and advice-giving roles within peer networks. These findings complement existing evidence that adolescent peer networks are often formed around gender \cite{boomgaarden_report_2022, djupe_present_2007, djupe_political_2018}, but go beyond prior work by showing how gender affects the flow of digital knowledge. Research has shown that the presence of more female peers in classrooms can improve academic and non-cognitive outcomes for all students \cite{gong2018effect, goulas2018does}, reinforcing the idea that diverse and inclusive peer environments matter. However, if digital skills are unevenly distributed across genders, homophily may exacerbate existing inequalities by limiting the sharing of important skills across genders.

In addition, our study illustrates how gendered self-perceptions influence peer learning dynamics. Girls, who tend to underestimate their digital skills \cite{haddon_childrens_2020}, are more likely to seek help, while boys, who often overestimate their skills \cite{haddon_childrens_2020}, are less likely to seek advice. These internalized perceptions shape how students navigate their learning networks and how others respond to them. Not only are girls more active as seekers and givers of advice, but they are also more likely to be seen as needing help, suggesting that peer responses may be based more on self-perceived than actual ability. Because self-assessed digital competence does not always match performance, the accuracy of peer recognition is uneven. Homophily amplifies these effects: when boys interact primarily with other boys, overconfidence may be reinforced, while girls in same-sex networks may further internalize underestimation of their skills.

This study makes several important contributions to research on digital skill development and adolescent peer learning. Most notably, it is the first to systematically examine how digital skills are shared within peer networks using sociocentric data from classroom settings. By moving beyond assumptions or qualitative observations, our network-analytic approach offers robust empirical evidence that informal peer dynamics -- who seeks advice from whom, and who is seen as a source of support -- play a central role in shaping digital skill acquisition. In doing so, we fill a significant gap in the digital literacy literature, which has traditionally focused on individual attributes or formal educational structures while overlooking the social contexts in which learning unfolds. Our findings extend this body of work by demonstrating how digital skill-sharing is structured by functional homophily and shaped by both objective and perceived competence, gendered patterns of interaction, and broader social dynamics.

These insights carry important implications for formal education. As our results show, peer networks serve as valuable ecosystems for collaborative learning, particularly when highly skilled students are positioned to share knowledge with others. This suggests that digital literacy programs should go beyond teacher-led instruction and actively incorporate peer learning strategies into classroom practices. Schools and curriculum designers should recognize and intentionally support these informal learning mechanisms, ensuring that they are inclusive and do not unintentionally reinforce existing inequalities. Here, the focus needs to be on bridging between skill levels and genders. On a broader policy level, our findings point to the importance of investing not only in formal instruction, but also in informal, peer-driven digital learning environments -- such as after-school clubs, mentoring programs, or online collaborative spaces -- that reflect how youth actually engage with technology. Recognizing the social dynamics of learning could help design more responsive, equitable, and sustainable approaches to digital education.

Gender-responsive design is particularly important. Our study shows that digital skill-sharing is shaped not only by competence but also by self-perception of competencies,  which are both structured by gender \cite{haddon_childrens_2020, mascheroni_explaining_2022}. Encouraging mixed-gender collaboration and creating environments where diverse learners can act as both knowledge seekers and providers can help break cycles of inequality. This is also relevant for parents and caregivers, whose role in digital education is often framed in regulatory terms. While guidance is essential, overly restrictive mediation can reduce opportunities for peer-based digital learning, especially for those who are already less confident in their abilities. A more empowering, dialogic form of mediation may provide greater support for adolescents' learning needs.

While this study offers novel insights, it is not without limitations. First, the data used here reflect a binary classification of gender, which limits our ability to speak to the experiences of non-binary and gender-diverse youth. Recent research has begun to explore how non-binary youth engage with digital technologies in different ways \cite{de-coninck_gendered_2023}, and future work should aim to include a broader range of gender identities. Second, while our cross-national dataset allows for generalizability within Western European settings, it does not allow for in-depth contextual analysis of country-specific patterns. More detailed analyses that explore cultural, institutional, or national variations would be a valuable next step. Methodologically, our study is limited by its reliance on cross-sectional, self-reported data. Future research would benefit from longitudinal and multi-method approaches that combine survey, behavioural, and observational data to better capture changes over time and validate reported behaviours. The inclusion of stable individual identifiers across waves would also improve the robustness of network analyses.

Future research should also address age-related variations in digital skills. Studies have shown that digital skills tend to increase with age across countries \cite{cabello-hutt_online_2018, kaarakainen_ict_2019, livingstone_risky_2013}, and it would be fruitful to examine how age influences peer network formation and skill-sharing practices. In addition, integrating academic achievement, cognitive skills, and attitudes toward technology into future models could provide a more complete picture of the conditions under which digital skills are developed. The context of peer interaction is another area for exploration: peer networks formed in schools may differ significantly from those formed outside formal educational settings, with different implications for behaviour and learning \cite{kiesner_peer_2003}.

Our study demonstrates the value of a sociocentric network approach to understanding how digital skills are acquired and shared within adolescent peer groups. By combining unique classroom network data with both objective and perceived measures of digital skills, we provide novel insights into how gender, homophily, and self-perception structure digital learning relationships. More broadly, this research highlights the importance of recognizing peer dynamics as a powerful force in shaping digital competence. Given that digital competence is increasingly linked to well-being \cite{livingstone_outcomes_2023}, educational success \cite{pagani_impact_2016}, labor market participation \cite{sanchez-canut_professional_2023}, and online safety \cite{quandt_dark_2022}, understanding the social dynamics of skill acquisition is more urgent than ever.

\section*{Funding}
This project has received funding from the European Union’s HORIZON2020 Research \& Innovation program under Grant Agreement no. 870612.

\newpage

\bibliographystyle{unsrt}  
\bibliography{references}  

\begin{appendices}

\newpage

\section{Additional Materials}\label{secA1}

\subsection{Example Skills Subdimension Questions}

All questions start with: \textit{Please indicate how true the following statements are of you when thinking about how you use the internet and technologies such as mobile phones or computers. Reply thinking about how true this would be of you if you had to do it now, on your own.}\newline

\noindent All items were measured using a 6-point Likert scale:

\begin{itemize}
    \item (0) I don't understand what you mean by this
    \item (1) not at all true of me
    \item (2) not very true of me
    \item (3) neither true nor untrue of me
    \item (4) mostly true of me
    \item (5) very true of me
\end{itemize}

\subsubsection*{Technical Skills}

\begin{itemize}
    \item I know how to turn off the location settings on mobile devices 
    \item I know how to store photos, documents or other files in the cloud (e.g. Google Drive, iCloud)
    \item I know how to use private browsing (e.g. incognito mode) 
\end{itemize}

\subsubsection*{Information Processing Skills}

\begin{itemize}
    \item I know how to choose the best keywords for online searches
    \item I know how to use advanced search functions in search engines 
    \item I know how to check if the information I find online is true 
\end{itemize}

\subsubsection*{Communication Skills}

\begin{itemize}
    \item Depending on the situation, I know which medium or tool to use to communicate with someone (e.g., make a call, send a WhatsApp message, send an email) 
    \item I know which images and information of me it is OK to share online  
    \item I know how to recognise when someone is being bullied online  
\end{itemize}

\subsubsection*{Content Creation Skills}

\begin{itemize}
    \item I know how to create something that combines different digital media (e.g., photo, music, videos, GIFs) 
    \item I know how to ensure that many people will see what I put online 
    \item I know how to distinguish sponsored and non-sponsored content online (e.g. in a video, in a social media post) 
\end{itemize}

\subsubsection*{Programming Skills}

\begin{itemize}
    \item I know how to use programming language (e.g. XML, Python)
\end{itemize}

\subsection{Pooled Posterior Estimates for ERGM terms}

\begin{sidewaystable}
\centering
\caption{ERGM Estimates for Research Questions and Hypotheses: Advice Seeking and Giving Networks\\
Posterior distributions of hierarchical bayesian models are provided; 95\% Credibility Intervals are in brackets.}
\label{tab:appendix}
\begin{tabular}{llcc}
\toprule
\textbf{Group} & \textbf{Parameter} & \textbf{Seeking} & \textbf{Giving} \\
\midrule
\multicolumn{4}{l}{\textbf{Research Questions}} \\
--   & Density                                & -4.12 [-4.25, -4.00]   & -3.99 [-4.11, -3.87]   \\
--   & Reciprocity                            & 2.84 [2.75, 2.93]      & 2.94 [2.86, 3.02]      \\
RQ1a & Sender effect of total skills          & -0.952 [-1.08, -0.826] & 0.163 [0.040, 0.285]   \\
RQ1a & Receiver effect of total skills        & 0.472 [0.346, 0.598]   & -0.338 [-0.460, -0.217]\\
RQ1b & Sender effect of perceived skills      & 0.119 [0.109, 0.130]   & 0.0705 [0.0606, 0.0801]\\
RQ1b & Receiver effect of perceived skills    & -0.0015 [-0.0110, 0.0083]& 0.0211 [0.0116, 0.0305]\\
RQ1c & Similarity in total skills (homophily) & -0.424 [-0.550, -0.300]& -0.369 [-0.486, -0.253]\\
RQ2  & Sender effect of being female          & 0.181 [0.125, 0.235]   & 0.0584 [0.0081, 0.110] \\
\cmidrule(lr){1-4}
\multicolumn{4}{l}{\textbf{Hypotheses}} \\
H1 & Density                                & -3.31 [-3.35, -3.27]  & -3.31 [-3.34, -3.27]  \\
H1 & Reciprocity                            & 2.77 [2.69, 2.85]     & 2.91 [2.83, 2.99]     \\
H1 & Receiver effect of being female        & 0.0681 [0.0194, 0.117]& 0.0563 [0.0096, 0.103]\\
\addlinespace
H2 & Density                                & -3.93 [-3.99, -3.87]  & -3.95 [-4.01, -3.90]  \\
H2 & Reciprocity                            & 2.51 [2.43, 2.61]     & 2.64 [2.55, 2.73]     \\
H2 & Same-gender preference (homophily)     & 1.05 [0.983, 1.11]    & 1.06 [1.00, 1.12]     \\
\bottomrule
\end{tabular}
\end{sidewaystable}

\end{appendices}

\end{document}